\documentclass{ws-procs975x65}

\begin{document}

\title{The soft X-ray excess AGN RE J2248-511}

\author{R.~L.~C. STARLING and E.~M. PUCHNAREWICZ}

\address{Mullard Space Science Laboratory, \\
Holmbury St. Mary, Dorking, Surrey, RH5 6NT, UK\\ 
E-mail: rlcs@mssl.ucl.ac.uk}

\author{E. ROMERO-COLMENERO}

\address{South African Astronomical Observatory, \\ 
P.O. Box 9, Observatory, 7935 South Africa\\
E-mail: erc@saao.ac.za}

\maketitle

\abstracts{We model the spectral energy distribution of the ultrasoft broad-line AGN RE J2248-511 with Comptonised accretion disc models. These are able to reproduce the steep optical and ultrasoft X-ray slopes, and the derived black hole mass is consistent with independent mass estimates. This AGN displays properties of both broad and narrow line Seyfert 1 galaxies, but we conclude that it is intrinsically a `normal' Seyfert 1 viewed at high inclination angle.}

\section{Introduction}
RE J2248-511 is a nearby ($z=0.101$) EUV-selected Seyfert galaxy discovered by the \emph{ROSAT}
Wide Field Camera\cite{Po}. Further observations showed this source to have a strong soft X-ray excess and spectral variability at optical\cite{Ma} \cite{Gru} and soft X-ray\cite{emp95} \cite{aad} wavelengths. However, none of these observations were simultaneous, so the existence of an optical to soft X-ray big blue bump (BBB) could not be confirmed. 

The soft X-ray
spectrum resembles those of narrow-line Seyfert 1 galaxies (NLS1's), but this AGN
has high velocity optical emission lines\cite{Ma} with H$\beta$ FWHM$\sim$2900 km s$^{-1}$, which classifies it as a Seyfert 1. Studies of \emph{ROSAT} PSPC slopes in AGN
had concluded that sources with both steep soft X-ray continuum slopes and broad optical emission lines are not found in nature\cite{BBF} \cite{WB} \cite{Gru}. This makes RE J2248-511 an unusual and ideal case in which to examine the relationship between the X-ray and optical continua, specifically the interpretation of the BBB as thermal emission from an accretion disc.
We describe the results of a multiwavelength monitoring
campaign of RE J2248-511, consisting of X-ray observations from the \emph{XMM} satellite with supporting quasi-simultaneous optical observations made at the South African Astronomical Observatory (SAAO) and archival multiwavelength
data. 

\section{Observations}
RE J2248-511 was observed by \emph{XMM-Newton}\cite{Jan} on 26 October 2000 and 31 October 2001, and we use the EPIC pn\cite{Stru} data here to investigate the broad-band X-ray spectrum. The resulting exposure times for the pn are 17.6 ks and 15.4 ks respectively, and no variability occurred during or between the observations.
The raw data from both observations were processed with the \emph{XMM} {\small SAS} v5.3. The 2 pn observations were coadded to form a single spectrum to increase signal to noise, and spectral analysis was done using {\small XSPEC} v11.2 in the energy range 0.3-10 keV. Optical spectra were taken in the same week as the first \emph{XMM} observation with the 1.9m at SAAO, and reduced using {\small IRAF}.

\section{The X-ray spectrum}
In one year, this source has shown low level flux variability, of order 10\%, and the hardness ratios are consistent with a constant value. A single absorbed\cite{emp95} power law is a poor fit to the data due to the presence of a strong soft excess ($\chi^{2}$/dof=1408/662). The power law is a good fit to the 2-10 keV range, and we find no evidence for a reflection component. The weakness of any Fe K$\alpha$ emission also suggests that there is no significant reflection in this spectrum. For the soft excess, 2 blackbodies or Comptonisation of soft photons in a hot plasma provides a good fit. The best fit to the 0.3-10 keV spectrum is a model of two blackbodies ($kT_{1}$=$0.09\pm0.01$ keV;$kT_{2}$=$0.21\pm0.03$ keV) plus a power law ($\Gamma$=1.8$\pm$0.08, $\chi^2$/dof=458/468). The X-ray flux is 1.16$\pm0.11$ $\times$ 10$^{-11}$ erg cm$^{-2}$ s$^{-1}$,
corresponding to a luminosity of 3.04$^{+0.26}_{-0.29}$ $\times$ 10$^{44}$ erg s$^{-1}$. 
Comparing the pn spectrum (by power law fitting) with data from previous X-ray missions, in their overlapping energy range
of 0.7-2 keV, the spectral slope during the \emph{XMM} and \emph{ROSAT} observations may be consistent, and are both in a softer state than during the \emph{ASCA} observation.

\begin{figure}
\begin{center}
\includegraphics[width=6cm, angle=90]{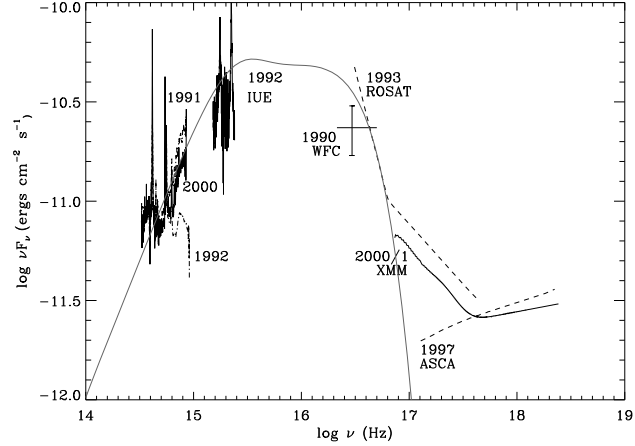}
\caption{All epoch spectral energy distribution, plus Comptonised accretion disc model as fitted to the October 2000 SED ($M=10^8$ M$_{\odot}$, $\dot{M}$=0.8$\dot{M}_{\rm Edd}$, face-on).}
\end{center}
\end{figure}

\section{The accretion disc and black hole mass}
The spectral energy distribution shows a strong big blue bump is present at most epochs (Fig 1.). We investigate the possibility of the entire optical to soft X-ray flux
originating in an accretion disc by applying Comptonised accretion disc models
developed by Czerny \& Elvis\cite{CE}, in which electron scattering occurs in the disc at high temperatures, to the spectral energy
distribution for October 2000. The variable parameters of the model are black hole mass, accretion rate and inclination. Inclination angles of $\cos{i}<0.75$ are ruled out since at low inclination angles (more edge-on discs) the ratio of optical to X-ray flux produced is too low. Black hole mass and accretion rate are more difficult to constrain since they are dependent upon each other. From this modelling the black hole mass range for these data is $10^{7.5}\le M \le 10^{8.5}$ M$_{\odot}$ and higher accretion rates $\ge 0.4 \dot{M}_{\rm Edd}$ are favoured. We can make an independent measure of black hole mass using the photoionisation method\cite{Kasp} \cite{WL}. We measure a velocity of the optical broad line region from our SAAO spectrum of $\sim$3000 km s$^{-1}$, which gives a black hole mass estimate of $M=10^8$ M$_{\odot}$, matching the modelling estimate. A lower limit on the black hole mass can be determined
by assuming that RE J2248-511 is not emitting at greater than the Eddington luminosity, and that the measured X-ray luminosity is 10\% of the bolometric luminosity\cite{Elv}. This gives $M\ge 2.3\times 10^7$ M$_{\odot}$.

\section{Discussion and conclusions}
We have shown that Comptonised accretion disc models, which treat the optical to soft X-ray emission as a single BBB, are able to comprise the majority of this flux. From these we derive constraints on the black hole mass which are fully consistent with the mass we obtain through the independent photoionisation technique. This implies that
thermal disc emission is the likely origin of the optical, UV and some of the soft X-ray continuum. An accretion disc does not, however, constitute the soft excess as observed with the pn, but it could be the origin of the `ultrasoft' component below 0.25 keV observed with \emph{ROSAT}.
The 0.3-2 keV soft component observed with \emph{XMM} has a blackbody-like shape, but is too hot to be blackbody disc emission. A model including Comptonisation of soft photons in a hot plasma provides a good fit. Between the \emph{XMM} observations, the X-ray spectral shape of RE J2248-511 remained approximately constant, while comparison with previous X-ray data shows long-term variability. This demonstrates that the `soft' state is a long-lived phase, and not, for example, a rapid flaring of the disc. 

Is RE J2248-511 an intermediate class of object linking the Seyfert 1's with the NLS1's, or a true Seyfert 1 seen with a particular observational bias? \\
Contrary to the proposed NLS1 scenario\cite{PDO},
the data are best fitted with high black hole masses ($\sim10^{8}$ M$_{\odot}$), whilst still favouring the high accretion rates suggested for NLS1 galaxies.
Therefore, it is not necessary for a black hole to have a low mass for the formation of an ultrasoft X-ray excess. Comparison of Comptonised accretion discs to the SED show that the orientation of the disc is close to face-on, allowing us to see a greater surface area of the accretion disc. If most of the EUV emission arises in the accretion disc then this source would appear EUV-bright compared with similar Seyfert galaxies viewed more edge-on. 
In the hard X-rays RE J2248-511 resembles a normal Seyfert 1 far more than a NLS1\cite{BME}. The soft X-ray flux measured with \emph{XMM} is typical of a Seyfert 1\cite{TP}, and its slope is similar to that found in the PG bright quasar sample \cite{Laor}. This source also follows the observed correlation between Balmer linewidth and soft X-ray slope\cite{emp92} \cite{Laor} if the `ultrasoft' part of the X-ray spectrum (below 0.25 keV) is excluded. Variability in the soft X-ray excess is a property which RE J2248-511 shares with the NLS1 galaxies, but changes are often more dramatic in NLS1's than observed here. 

The strength and broad wavelength span of
the BBB in RE J2248-511 is unusual for a broad-line AGN, but we find this may be explained by a face-on Comptonised
accretion disc, likely to be accreting at a high rate onto a 10$^{7.5}$-10$^{8.5}$ M$_{\odot}$ black hole. We propose that this source is intrinsically a normal Seyfert 1, but shares the ultrasoft X-ray excess property with the NLS1's because it is observed at a higher inclination angle than the majority of Seyfert 1 galaxies.

\section*{Acknowledgments}
RLCS acknowledges financial support from a PPARC studentship.
%
%
%
%

\end{document}